\documentstyle[aps,prl,epsfig,multicol]{revtex}
\tighten
\begin{document}
\draft

\title{Zero-bias anomaly in disordered wires}
\author{E.G. Mishchenko$^{1,2,3}$, A.V. Andreev$^{1,2}$ and L.I.
  Glazman$^{4}$} 
\address{{}$^1$Bell Laboratories, Lucent Technologies, Murray Hill, NJ 07974}
\address{{}$^2$Department of Physics, University of
  Colorado, CB 390, Boulder, CO 80309-0390} \address{{}$^3$ L. D. Landau
  Institute for Theoretical Physics, Russian Academy of Sciences,
  Moscow 117334, Russia} \address{{}$^4$ Theoretical Physics
  Institute, University of Minnesota, Minneapolis, MN 55455}
  \maketitle

\begin{abstract}
  We calculate the low-energy tunneling density of states
  $\nu(\epsilon, T)$ of an $N$-channel disordered wire, taking into
  account the electron-electron interaction non-perturbatively. The
  finite scattering rate $1/\tau$ results in a crossover from the
  Luttinger liquid behavior at higher energies,
  $\nu\propto\epsilon^\alpha$, to the exponential dependence $\nu
  (\epsilon, T=0)\propto \exp{(-\epsilon^*/\epsilon)}$ at low
  energies, where $\epsilon^*\propto 1/(N \tau)$.  At finite
  temperature $T$, the tunneling density of states depends on the
  energy through the dimensionless variable $\epsilon/\sqrt{\epsilon^*
  T}$. At the Fermi level $\nu(\epsilon=0,T) \propto \exp (-\sqrt{\epsilon^*/
  T})$.
\end{abstract} \pacs{PACS numbers: 73.63.-b, 73.23.Hk, 73.21.Hb}

\begin{multicols}{2} 
  
The influence of electron-electron interactions on transport in
disordered systems has been extensively investigated for the past two
decades~\cite{AA}. It is well known that the interaction has the
strongest effect in low-dimensional systems. Electron tunneling into a
one-dimesional conductor is suppressed by interactions even in the
absense of disorder. This suppression, which yields vanishing
tunneling density of states (TDOS) at the Fermi level, can be
described in the framework of the Luttinger liquid theory. The
recently discovered carbon nanotubes provide a unique opportunity for
studying interaction effects in quantum wires\cite{Ta,Bo,Ko,Ba}.  
Although the
properties of single-wall nanotubes are well described by the
Luttinger liquid theory\cite{YPB}, the transport in multi-wall
nanotubes (MWNT) is still not very well understood.  The number of
channels, disorder strength, and carrier concentrations in these
systems can vary over a wide range and are difficult to control
experimentally.  Whereas some measurements indicate ballistic electron
transport~\cite{F}, most experiments exhibit diffusive electron motion
\cite{Ba,Ba1,La,Sch}.  Furthermore, experiments\cite{Ba2} demonstrate a
strong suppression of TDOS $\nu (\epsilon)$ near the Fermi level
($\epsilon =0$). On the other hand, the existing microscopic
theory\cite{AA} treats the screened Coulomb interaction in the first
order of perturbation theory; it provides the result for the
correction to the density of states, $\delta\nu(\epsilon)\propto
-1/\sqrt{\epsilon}$, which is valid as long as $\delta\nu(\epsilon)$
is small. Clearly, this result of the lowest-order perturbation theory
is insufficient for finding the behavior of TDOS $\nu (\epsilon)$ in
the limit of $\epsilon\to 0$. 

In this paper we present a theory of the zero-bias anomaly in the
tunneling density of states in quantum wires. We
treat the dynamically screened Coulomb interaction non-perturbatively
and allow for an arbitrary value of $\epsilon\tau$, where $\tau$ is the elastic
momentum relaxation time of electrons, and energy $\epsilon$ is
measured from the Fermi level. This enables us to describe the
crossover from the known Luttinger liquid results\cite{MG} valid at
higher energies, $\epsilon \tau \gg 1/\sqrt{N}$, to the new low-energy
($\epsilon \tau \ll 1/\sqrt{N}$) behavior of the TDOS: 
\begin{equation}
\nu(\epsilon,T)\propto \exp\left\{-\sqrt\frac{\epsilon^*}{T}
F\left(\frac{\epsilon}{\sqrt{\epsilon^*T}}\right)\right\}
\label{zba}
\end{equation}
(hereinafter we use units with $\hbar=k_B=1$). The characteristic
energy $\epsilon^*$ here depends on the interaction strength $g$,
\begin{equation}
\label{g2}
\epsilon^*=\frac{g}{\pi N\tau},\quad g = \frac{\pi e^2}{4 \bar{v}}
\ln{\frac{d}{R}}\,\, , 
\end{equation}
on the number of channels $N$ in the quantum wire, and on $\tau$. Here
$\bar{v}$ is the Fermi velocity averaged over all channels, and $d\gg
R$ is the distance at which the electric field is shielded\cite{Zuzin}
say, by conducting electrodes surrounding the wire. The scaling
function $F(x)$ and its asymptotics at $x \ll 1$ (regime considered
first in Ref.~\cite{Naz}) and $x \gg 1$, are presented in Fig.~1. The
result (\ref{zba}) applies at sufficiently low temperatures and
energies, when the value of the exponent in this equation is large.
Note that according to Eq.~(\ref{zba}), at finite $T$ the
characteristic scale for the energy-dependence of TDOS is given not by
$T$, but by a much larger value $\sqrt{\epsilon^*T}$.
\begin{figure}
\begin{center}
 \epsfxsize=7.5cm \epsfbox{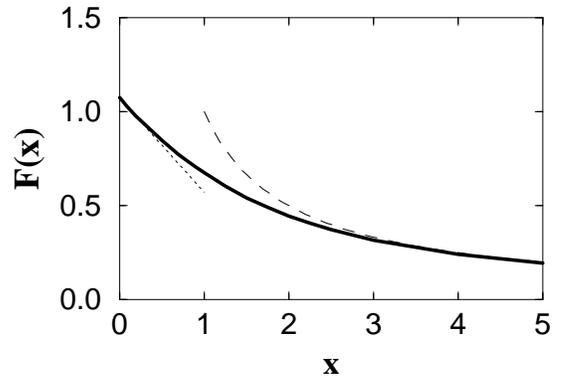}
\end{center}
\caption{ The scaling function $F(x)$ and its asymptotics:
$F(x)=1.07-x/2$ for $x \ll 1$ (dotted line), and $F(x) \sim 1/x$
for $x \gg 1$ (dashed line).}
\label{fig:1}
\end{figure} 
In the absence of shielding Eq.~(\ref{zba}) is somewhat modified, as
the interaction parameter $g$ becomes a weak function of temperature
and energy. We briefly discuss this case in the end of the paper,
see Eqs.~(\ref{newg})--(\ref{ln2}).

Having in mind experimental applications, we focus our discussion on
carbon nanotubes.  A typical MWNT consists of several (up to ten)
graphite monolayer sheets rolled concentrically into cylinders. At
zero doping they can be either metallic or semiconducting, depending
on the helical arrangement of the carbon hexagons.  The electron band
structure of a single carbon nanotube~\cite{bandstruct} has two points
in the Brillouin zone with the Dirac-like spectrum $\epsilon_k =
v(k^2+k_{\perp}^2)^{1/2}$; the velocity here is $v \simeq 8 \cdot
10^7$ cm/s, and the transverse momentum $k_{\perp}=n/R$ is quantized
due to the periodic boundary conditions around the circumference,
$-k_F R \le n \le k_F R$.  The number of conducting sub-bands around
each Dirac point $N=2k_F R$ is determined by the radius of the
nanotube and by the doping level, $\mu=vk_F$ ($\sim 0.5$~eV).  A
typical radius of the outermost shell is of the order of $10$~nm.
Each sub-band has its own Fermi velocity $v_n=v\sqrt{1-n^2/k_F^2R^2}$
and momentum $k_F v_n/v$ along the cylinder axis. The electrons are
scattered between different sub-bands within the same tube by
impurities, lattice imperfections and by the incommensurate lattice
potential of the neighbouring tubes.  We consider the experimentally
relevant case of the ballistic electron motion around the circumference,
$l=v\tau > R$, and concentrate on low-energy, $\epsilon < v/R$, limit.
Electron tunneling at higher energies, $\epsilon > vl/R^2$, in the
opposite case of a short mean free path, $l<R$, was recently discussed in
Ref.~\cite{EG}.

In the measurements of the TDOS the tunneling current
propagates through the outermost shell\cite{Ba1} while the inter-shell
tunneling is largely suppressed. The role of the electrons in inner
shells is then believed to be reduced merely to the dynamical
screening of the Coulomb interaction between the electrons in the
outer shell. Little is known about the contribution of inner-shell
electrons to screening since the doping level in the inner shells is
 difficult to characterize experimentally. Two distinct scenarios
can be imagined: i) the dopants are outside the nanotube, and the
doping electrons reside in the outer shell only. ii) the dopants are
distributed uniformly inside the MWNT, which leads to a uniform
density of carriers across the shells. Below we concentrate on the
first scenario, when the inner shells of the tube may be ignored. The
second scenario will be considered elsewhere.

We start with calculating the TDOS in the first order in the screened
interaction potential. This calculation follows the well-known route
first developed for the case of a diffusive electron motion\cite{AA79},
and extended later~\cite{RAG,KhR} to the case of an arbitrary value of
$\epsilon\tau$. The zero-temperature result can be cast in the
familiar\cite{AA79,AAL} form,
\begin{mathletters}
\begin{eqnarray}
\label{correction}
\frac{\delta \nu (\epsilon)}{\nu_0}&=& \int_{\epsilon}^{\infty}
{d\omega}~
{\cal V}(\omega),\\
\label{v}
{\cal V}(\omega)&=&
\Im \sum_{q_{\perp}}
\int\limits_{-\infty}^{\infty} \frac{dq}{2\pi^2} \Gamma^2(\omega,{\bf
  q}) {\cal G}^2 (\omega,{\bf q})U(\omega, {\bf q}).
\end{eqnarray}
\end{mathletters}
Here $\Gamma$ is the impurity-renormalized vertex. Its inverse is
given by the usual impurity ladder,
\begin{eqnarray}
\label{diffuson}
 \Gamma^{-1}(\omega,{\bf q})&\equiv&\langle \frac{\omega-{\bf qv}}{\omega-{\bf qv}
   +i/\tau}\rangle 
\nonumber\\&=&
\frac{1}{\pi\nu_0}
\sum_n  
v_n^{-1} \frac{\omega(\omega+i/\tau)-(qv_n+q_{\perp}v_{\perp n})^2}
{(\omega+i/\tau)^2-(qv_n+q_{\perp}v_{\perp n})^2}.
\end{eqnarray}
The product of Green functions averaged over the Fermi surface
[abbreviated as ${\cal G}^2$ in Eq.~(\ref{correction})] equals
\begin{eqnarray}
\label{produ}
{\cal G}^2(\omega,{\bf q}) &\equiv&\langle \frac{1}{(\omega-{\bf qv} +i/\tau)^2}\rangle
\nonumber\\ 
&=&
\frac{1}{\pi\nu_0}
\sum_n 
v_n^{-1} \frac{(\omega+i/\tau)^2+(qv_n+q_{\perp}v_{\perp n})^2}
{[(\omega+i/\tau)^2-(qv_n+q_{\perp}v_{\perp n})^2]^2},
\end{eqnarray}
where ${\bf qv}= qv_n+q_{\perp} v_{\perp n}$ and $v_{\perp n}=vn/(k_F
R)$ is the transverse velocity in the $n$-th band, $\nu_0=\sum_n (\pi
v_n)^{-1}$ is the total density of states in the outermost shell (the
summation  accounts also for both spin directions and the
presence of two Dirac points in the Brillouin zone).

The function ${U}(\omega, {\bf q})$ in Eq.~(\ref{v}) represents the
dynamically screened Coulomb interaction of electrons and is given by
\begin{equation}
\label{Coulomb}
U(\omega,{\bf q}) = \frac{V({\bf q})}{1-
V({\bf q}) ~\Pi (\omega,{\bf q})}.
\end{equation}
Within the assumptions of our model, the polarization operator here,
\begin{equation}
\label{pol}
\Pi(\omega,q,q_m)= \nu_0 \Gamma (\omega,{\bf q})~
\langle \frac{{\bf qv}}{\omega-{\bf qv} +i/\tau} \rangle,
\end{equation}
is provided by the outer-shell electrons. The bare Coulomb
potential in Eq.~(\ref{Coulomb}) is
\begin{equation}
\label{bare}
V (q,q_m)=
\frac{2e^2}{\pi}\int_0^{\pi} d\phi 
K_0 \left(2qR \sin{\frac{\phi}{2}}\right)\cos{m\phi},
\end{equation}
where $K_0(x)$ is the modified Bessel function, and we used the fact
that the momenta along the circumference of the tube are quantized and
given by $q_m=m/R$.

To consider the low-energy behavior of $\delta\nu (\epsilon)$, we will
need only the long-range limit, $qR \ll 1$, of
Eqs.~(\ref{Coulomb})--(\ref{bare}); we find for the bare interaction
\begin{eqnarray}
  \label{vij}
V(q,q_m)=\begin{cases}{
\frac{e^2}{\vert m \vert}, & $m \neq 0$, \cr 
e^2  \ln{[\min{(\frac{d^2}{R^2}, \frac{1}{q^2 R^2})}]},& $m = 0$.}
\end{cases}
\end{eqnarray}
The integrals in Eqs.~(\ref{correction}) and (\ref{v}) are dominated by
the region of high frequencies $\omega \gg qv$, where the dynamically
screened Coulomb interaction (\ref{Coulomb}) has plasmon poles, and
can be rewritten as
\begin{eqnarray}
\label{correction2}
{\cal V}(\omega) =
\Im \sum_{m} \int_{-\infty}^{\infty} \frac{dq}{2\pi^2}  
 \frac{ (\omega+i/\tau) V(q,q_m)}{\omega\left[\omega
     (\omega+i/\tau)- \omega_m^2(q)\right]}.
\end{eqnarray}
Here $\omega_m(q)$ denotes the frequency of the plasmon excitations
which in the long wave length limit, $q d \ll 1$, is given by the
following equation:
\begin{equation}
\label{plasmon}
\omega^2_m(q)
=  \begin{cases}{
e^2\nu_0  q^2 v_{\parallel}^2  
\ln{[\min{(\frac{d^2}{R^2}, \frac{1}{q^2 R^2})}]} 
, & $m=0$, \cr
\frac{e^2\nu_0}{\vert m \vert} \left( q^2 v_{\parallel}^2 +\frac{
    v_{\perp}^2 m^2}{ R^2}\right), & $m \ne 0$.}
\end{cases}
\end{equation}
In Eq.\ (\ref{plasmon}) we introduced the average squares of the
longitudinal, $ v_{\parallel}^2$, and transverse, $v_{\perp}^2$,
electron velocities
$$v_{\parallel}^2=\frac{\sum_n v_n }{\sum_n v_n^{-1}}, ~~~~~
v_{\perp}^2=\frac{ v^2 \sum_n v_n^{-1} n^2} { (k_F R)^2\sum_n v_n^{-1}}.$$

For $\epsilon > v_\perp/R$ in Eq.~(\ref{correction}),  
we can approximate the sum over $m$ 
in Eq.~(\ref{correction2}) by an integral
recovering the ballistic counterpart ~\cite{KhR} of the two-dimensional
diffusive correction discussed by Egger and Gogolin~\cite{EG} 
for short-range 
interaction. 
However, in contrast to their conclusions, for
lower energies $\epsilon < v_\perp/R$ the contribution of $m\neq 0$
terms becomes energy independent for both short-range and
Coulomb interaction, in the latter case because of the gaps in plasmon
spectra. The $m=0$ plasmon, on the other hand, is gapless. Its
contribution to Eq.~(\ref{correction2}) depends on $\epsilon$, and has
a singularity at $\epsilon \to 0$.  Therefore, to study the energy
dependence of DOS at low energies we neglect the non-singular
contribution of the $m\neq 0$ modes and retain only the $m=0$ term
in Eq.~(\ref{correction2}).  The expression (\ref{plasmon}) for $m=0$
ceases to be correct at frequencies larger that $\sqrt{N} \bar{v}/R$
which correspond to plasmons with wavelength of order of the tube
radius $R$, representing the obvious ultraviolet cut-off for
one-dimensional effects.  Performing the integral over the momenta in
Eq.~(\ref{correction2}), we obtain
\begin{equation}
\label{v1}
{\cal V}(\omega) = - \left(\frac{e^2
\ln{[\min{(\case{d}{R}, \case{\sqrt{N} v}{R\omega})}]}}{2\pi N \bar{v}}
\right)^{1/2}
\Re 
~ \frac{\sqrt{\omega +\frac{i}{\tau}}}{\omega ^{3/2}},
\end{equation}
where $\bar v=\sum_nv_n/N$ is the average Fermi velocity.  The two
distinct regions of the frequency dependence here, $\omega\gg 1/\tau$
and $\omega\ll 1/\tau$, define two domains for the energy-dependent
correction to the TDOS.  Substituting
${\cal V}(\omega)$ into Eq.~(\ref{correction}), 
and assuming that $\bar v \tau > d/\sqrt{N}$, we obtain
\begin{equation}
\label{pertur}
\frac{\delta \nu (\epsilon)}{\nu_0} =
-\frac{1}{\pi}\sqrt{\frac{2g}{N}}
\begin{cases}{
\sqrt{\frac{2}{\epsilon\tau}}
+
\ln{\lambda \tau}, &
$\epsilon < 1/\tau,$ \cr
\ln{(\lambda/\epsilon)}, &
 $\epsilon > 
 1/\tau,$ }\end{cases}  
\end{equation}
where $g$ is
defined in Eq.~(\ref{g2}), and 
$\lambda=\bar v\sqrt{N}/(R^2d)^{1/3}$.
The first term of the low-energy asymptotic here is familiar from the
zero-bias anomaly theory of Altshuler and Aronov\cite{AA}, while the
second term represents the omitted in\cite{AA} high-frequency
($\omega\gg1/\tau$) contribution to the integral
Eq.~(\ref{correction}). The behavior of $\delta\nu$ at $\epsilon\gg
1/\tau$ corresponds to the ballistic electron motion.

The perturbative expressions diverge at the Fermi level ($\epsilon \to
0$). To calculate the tunneling DOS to all orders in the interaction constant at
small $\epsilon$, we make use of the phase approximation for the
fluctuating potential induced by the electron-electron interaction
\cite{Naz}.  The non-perturbative expression for the density of states
can be cast in the form equivalent to the one derived in Ref.~\cite{KA},
\begin{eqnarray}
\label{dosexp}
\frac{\nu (\epsilon,T)}{\nu_0} = T \cosh{\frac{\epsilon}{2T}}
\int_{-\infty}^\infty dt~ \frac{\cos{\epsilon t}}{\cosh{\pi T t }}~
\nonumber\\ 
\times
 \exp{\left\{ \int_0^{\infty} d\omega~
{\cal V} (\omega)~ \frac{\cosh{\case{\omega}{2T}} - \cos{\omega t}}
{\sinh{\case{\omega}{2T}}} \right\} }.
\end{eqnarray}
In the region of validity of Eq.~(\ref{dosexp}), the time integral
can be evaluated within the saddle-point approximation. As the
saddle point lies on the imaginary axis in the interval $0 \le -i t <
1/2T$, the denominator in Eq.\ (\ref{dosexp}) is always a slowly
varying function and does not contribute to the saddle point exponent.

We consider first the ballistic regime, $\epsilon >\sqrt{\epsilon^*/\tau}$.
It allows us to reproduce the conventional Luttinger liquid results,
so for brevity we mention here only the limit $T\to 0$.
Evaluating the integrand of Eq.~(\ref{dosexp}), we find
\begin{equation}
\label{dosball}
\nu (\epsilon)\sim 
\nu_0\left(\frac{\epsilon}{\lambda \sqrt{\epsilon^* \tau }}
\right)^\alpha;\quad 
\alpha =\frac{1}{\pi} \sqrt{\frac{2g}{N}}.
\end{equation}

In the diffusive regime, $\epsilon < \sqrt{\epsilon^*/\tau}$, the main
contribution to the TDOS comes from the small frequencies, where one can
reduce Eq.~(\ref{v1}) to ${\cal V}(\omega) =\sqrt{\epsilon^*/\pi
\omega^{3}}$. Furthermore, if $\epsilon\ll 1/N\tau$ and $T\ll g/\pi
N\tau$,  the saddle point approximation is again applicable for the evaluation
of the time integral in Eq.~(\ref{dosexp}). One can easily check that at
the saddle point the $\cosh{\pi Tt}$ function in Eq.~(\ref{dosexp})
can be replaced by $1$, and therefore the density of states satisfies
the scaling form,
\begin{equation}
\label{scaling}
{\nu(\epsilon, T)} \sim \frac{\nu_0}{(\lambda \tau)^\alpha}
\exp \left\{-\sqrt{\frac{\epsilon^*}{T}}
F\left(\frac{\epsilon}{\sqrt{\epsilon^*T}}\right)\right\}.
\end{equation}
Here the function
\begin{equation}
\label{fx}
F(x)=\int_0^{\infty}\!\!\! dy\frac{\cosh{\case{y}{2}} - \cosh{yz_s(x)}}
{\sqrt{\pi}~y^{3/2}\sinh{\frac{y}{2}}} -
x\left[\case{1}{2}-z_s(x)\right]
\end{equation}
is determined by the value of the integrand in Eq.~(\ref{dosexp}) at
the saddle point $t_s\equiv iz_s/T$. The dependence of $z_s$ on the
ratio $\epsilon/\sqrt{\epsilon^*T}\equiv x$ is given by equation:
\begin{equation}
\label{saddle}
x=\frac{1}{\sqrt{\pi}}
 \int_0^\infty \frac{ dy~\sinh{z_s y}}{\sqrt{y}\sinh{y/2}}.
\end{equation}
Numerical solution of this parameter-free equation and the subsequent
evaluation of the integral in Eq.~(\ref{fx}) yields the graph of
scaling function $F(x)$ plotted in Fig.~1 and its asymptotics given in the
figure caption. Note that the pre-exponential factor in
Eq.~(\ref{scaling}), which we omitted in Eq.~(\ref{zba}) for the sake
of brevity, provides the proper matching of the results
(\ref{dosball}) and (\ref{scaling}) obtained in the energy domains
$\epsilon\gg\sqrt{\epsilon^*/\tau}$ and $\epsilon\ll
\sqrt{\epsilon^*/\tau}$, respectively.

We derived our main results, Eqs.~(\ref{scaling})--(\ref{saddle}),
assuming that the Coulomb interaction is shielded at distances $\sim
d$. In the absence of shielding, the long-range nature of the
interaction potential leads to a stronger\cite{GRS} than predicted by
Eq.~(\ref{dosball}) suppression of the TDOS in the ballistic regime,
$\ln [\nu (\epsilon)/\nu_0]\propto -[\ln
(\epsilon/\epsilon^*)]^{3/2}$.  At lower energies (corresponding to
the diffusive regime), the effect of the long-range potential can be
accounted for by replacing the parameter $\epsilon^*$ of
Eq.~(\ref{g2}) with the logarithmic function of energy and
temperature,
\begin{equation}
\epsilon^*\to \epsilon^*(\epsilon,T)=\epsilon_0
\ln{\frac{\bar{v}/R}{\mbox{max}(\epsilon,\sqrt{T\epsilon_0})}},\quad
\epsilon_0\equiv\frac{e^2}{4\bar{v}N\tau}.
\label{newg}
\end{equation}
After the definition of $\epsilon^*$ is adjusted to reflect this
replacement, we can use Eqs.~(\ref{scaling})--(\ref{saddle}) again. 
In the limit of low temperatures, $T\ll\epsilon^2/\epsilon_0$, we find:
\begin{equation}
\label{ln1}
\nu(\epsilon)\propto
\exp\left(-\frac{\epsilon_0}{\epsilon}\ln\frac{\bar{v}}{\epsilon
    R}\right).
\end{equation}
The suppression of the TDOS near the Fermi surface
($\epsilon\ll\sqrt{T\epsilon_0}$) at finite temperatures is given by
\begin{equation}
\label{ln2}
\nu(T)\propto
\exp\left[-1.07
\sqrt{\frac{\epsilon_0}{T}} 
\ln^{1/2}\left(\frac{\bar{v}}{R\sqrt{T\epsilon_0}}\right)
\right].
\end{equation}

The origin of the strong suppression of the TDOS, see
Eqs.~(\ref{dosball}), (\ref{scaling}), (\ref{ln1}), and (\ref{ln2}),
lies in the re-distribution of charge of the tunneling electron along
the wire.  This process is impeded by the finite propagation time of
plasmons which enable the charge spreading. The requirement that the
wavelength of the relevant plasmons $\bar{v} \sqrt{N}/\epsilon$ be
shorter that the length of the wire $L$ imposes the lower energy limit
for the applicability of the present theory.  For smaller energies,
$\epsilon < \bar{v} \sqrt{N} /L$, the tunneling DOS of the wire
depends on the impedance of the leads attached to the segment\cite{Z}.
To observe a sizable suppression of TDOS, the nanotube must be
sufficiently long, and therefore have high intrinsic resistance. We
find that in order to reach the strong diffusive renormalization of
TDOS anomaly ($\epsilon,T\ll\epsilon^*$) the total resistance of the
segment should be made larger than $(h/e^2)\sqrt{N/g}$.  This
condition on the overall segment length does not invalidate the
employed method which assumes the diffusive motion of
electrons. Indeed, the characteristic frequencies of the plasmons
involved are high enough for the weak localization's corrections to be
ignored in the calculation of TDOS. In order to avoid localization
effects in the DC transport measurement, the two junctions needed for
the TDOS measurement should be attached to the MWNT within a distance
shorter than the localization length from each other.

We presented above the derivation of TDOS only for the case of a
relatively long mean free path $l$, exceeding the radius $R$ of a MWNT,
but our main result~(\ref{zba}) is valid for any
relation between $l$ and $R$. The only additional feature appearing in
the case $l\ll R$, is the studied in Ref.~\cite{EG} intermediate range of
energies, where the TDOS behaves as in a two-dimensional disordered
conductor.

In summary, we have obtained the tunneling density of states in a
disordered quasi-one-dimensional conductor. Our results are
non-perturative in the electron-electron interaction, and cover both
the diffusive and ballistic regimes of the electron motion. 
In contrast to the two-dimensional
case~\cite{KA,Finkelstein}, the non-perturbative results
(\ref{dosexp}) and (\ref{ln1}) are not given by a simple
exponentiation of the first-order interaction correction
(\ref{pertur}). 

We thank B.~ Altshuler, I.~Beloborodov, A.~Kamenev, P.~McEuen and M.~Paalanen 
for discussions. This research
was sponsored by the NSF Grants DMR-9731756, DMR-9984002 and
BSF-9800338 and by the A.P.~Sloan and the Packard Foundations.  E.~M. 
acknowledges the support of the Russian Foundation for Basic Research,
Grant 01-02-16211.

\end{multicols}
\end{document}